\documentclass[prb,aps,showpacs,twocolumn,superscriptaddress,nobibnotes,epsf]{revtex4}
\usepackage{graphicx}
\usepackage{dcolumn}
\usepackage{bm}
\usepackage{tabularx}
\usepackage{amsmath}
\usepackage{textcomp}
\usepackage{upgreek}

\begin{document}
\title{Multi-gap nodeless superconductivity in nickel chalcogenide TlNi$_2$Se$_2$}

\author{X. C. Hong}
\affiliation{State Key Laboratory of Surface Physics, Department of Physics, and Laboratory of Advanced Materials, Fudan University,
Shanghai 200433, P. R. China}

\author{Z. Zhang}
\affiliation{State Key Laboratory of Surface Physics, Department of Physics, and Laboratory of Advanced Materials, Fudan University,
Shanghai 200433, P. R. China}

\author{S. Y. Zhou}
\affiliation{State Key Laboratory of Surface Physics, Department of Physics, and Laboratory of Advanced Materials, Fudan University,
Shanghai 200433, P. R. China}

\author{J. Pan}
\affiliation{State Key Laboratory of Surface Physics, Department of Physics, and Laboratory of Advanced Materials, Fudan University,
Shanghai 200433, P. R. China}

\author{Y. Xu}
\affiliation{State Key Laboratory of Surface Physics, Department of Physics, and Laboratory of Advanced Materials, Fudan University,
Shanghai 200433, P. R. China}

\author{Hangdong Wang}
\affiliation{Department of Physics, Zhejiang University, Hangzhou 310027, P. R. China}
\affiliation{Department of Physics, Hangzhou Normal University, Hangzhou 310036, P. R. China}

\author{Qianhui Mao}
\affiliation{Department of Physics, Zhejiang University, Hangzhou 310027, P. R. China}

\author{Minghu Fang}
\affiliation{Department of Physics, Zhejiang University, Hangzhou 310027, P. R. China}

\author{J. K. Dong}
\email{jkdong@fudan.edu.cn}
\affiliation{State Key Laboratory of Surface Physics, Department of Physics, and Laboratory of Advanced Materials, Fudan University,
Shanghai 200433, P. R. China}

\author{S. Y. Li}
\email{shiyan_li@fudan.edu.cn}
\affiliation{State Key Laboratory of Surface Physics, Department of Physics, and Laboratory of Advanced Materials, Fudan University,
Shanghai 200433, P. R. China}

\date{\today}

\begin{abstract}
Low-temperature thermal conductivity measurements were performed on single crystals of TlNi$_2$Se$_2$, a nickel-chalcogenide heavy-electron superconductor with $T_c$ $\simeq$ 3.7 K. In zero field, the residual electronic contribution at $T$ $\rightarrow$ 0 K ($\kappa_0/T$) was well separated from the total thermal conductivity, which is less than 0.45\% of its normal-state value. Such a tiny residual $\kappa_0/T$ is unlikely contributed by the nodal quasiparticles.
Nodeless gap structure is supported by the very weak field dependence of $\kappa_0(H)/T$ in low magnetic fields. In the whole field range, $\kappa_0(H)/T$ exhibits an ``$S$''-shape curve, as in the case of nickel pnictides BaNi$_2$As$_2$ and SrNi$_2$P$_2$. This common feature of nickel-based superconductors can be explained by multiple nodeless superconducting gaps.

\end{abstract}


\maketitle
\section{Introduction}

The iron-based high-$T_c$ superconductors (IBSs) can be classified into two main groups: iron pnictides and iron chalcogenides. \cite{review-AdvPhys,review-NatPhys,review-RMP}
Following the IBSs, it was found that most of their nickel-based counterparts with the same crystal structure are also superconducting. \cite{NBS1,NBS2,NBS3,NBS4}
The nickel-based superconductors (NBSs) exhibit some distinct features: (i) the superconducting transition temperature $T_c$ is pretty low, usually lower than 5 K; \cite{RonningF} (ii) the Fermi surfaces are more complicated and three-dimensional; \cite{Terashima,Subedi,ZhouB} (iii) there is no evidence for the existence of an antiferromagnetic order neighboring the superconducting state.

It is important to know whether the pairing mechanism of NBSs is different from IBSs. Clarifying the superconducting gap symmetry and structure will provide important clues. For the IBSs, it has been shown that the gap structure is quite elusive, varying substantially from member to member and as a function of doping. \cite{RPP} While most IBSs have multiple nodeless gaps (likely $s_\pm$-wave), some of them show nodal superconductivity. \cite{RPP} In the case of NBSs, however, the superconducting gap appears more ``conventional". Both specific heat and thermal conductivity measurements suggest fully gapped $s$-wave superconductivity in BaNi$_2$As$_2$. \cite{Kurita} Isovalent phosphorus doping does not change its gap structure, \cite{BaNi2AsP} in contrast to that observed in nodal superconductor BaFe$_2$(As$_{1-x}$P$_x$)$_2$. \cite{Hashimoto} Low-temperature magnetothermal conductivity $\kappa(T,H)$ measurements also rule out the presence of nodes in the superconducting gap of SrNi$_2$P$_2$. \cite{NKurita}

While fully gapped $s$-wave superconductivity seems to be a universal feature of nickel pnictides, it is not so clear for nickel chalcogenides, in which heavy-electron behavior was observed. \cite{Neilson,R.Neilson,FangMH} In the three nickel chalcogenides, KNi$_2$Se$_2$ ($T_c$ $\simeq$ 0.80 K), KNi$_2$S$_2$ ($T_c$ $\simeq$ 0.46 K), and TlNi$_2$Se$_2$ ($T_c$ $\simeq$ 3.7 K), the electronic specific-heat coefficient $\gamma$ is 44, 68, and 40 mJ mol$^{-1}$ K$^{-2}$, respectively. \cite{Neilson,R.Neilson,FangMH} The estimated effective electron mass $m^{\ast}$ can be as high as 24$m_e$ in KNi$_2$S$_2$. \cite{R.Neilson} The $\gamma(H)$ of TlNi$_2$Se$_2$ exhibits a square root field dependence, which is usually seen in nodal superconductors.\cite{FangMH} Therefore, it is of great interests to investigate the superconducting gap structure of nickel chalcogenides. For KNi$_2$Se$_2$ and KNi$_2$S$_2$, only polycrystalline samples were synthesized so far. Fortunately, sizable high-quality single crystals of TlNi$_2$Se$_2$ have been successfully grown. \cite{FangMH}

Low-temperature heat transport is an established bulk technique to study the superconducting gap structure.\cite{Louis}
In this paper, we present the thermal conductivity measurements of TlNi$_2$Se$_2$ single crystals down to 50 mK ($\sim$ $T_c$/70). The relatively tiny residual linear term in zero field and its slow field dependence in low fields suggest nodeless superconducting gap. In the whole field range, the $\kappa_0(H)/T$ curve shows a concave to convex evolution (``$S$'' shape), which was previously also observed in BaNi$_2$As$_2$ and SrNi$_2$P$_2$. Multi-gap superconductivity is introduced to explain this common behavior of NBSs.

\section{Experiment}

Single crystals of TlNi$_2$Se$_2$ were grown using the self-flux method. \cite{FangMH} The dc magnetization was measured by a superconducting quantum interference device (MPMS, Quantum Design). Two samples, labeled as A and B, were used in the transport measurements. The two samples were cleaved to rectangular shape with dimensions of $\sim$2.0 $\times$ 0.5 mm$^2$ in the \emph{ab} plane and $\sim$40 $\mu$m along the \emph{c} axis. Contacts were made directly on the sample surfaces with silver paint, which were used for both resistivity and thermal conductivity measurements. The contacts are metallic with typical resistance of 10 m$\Omega$ at 2 K. In-plane thermal conductivity was measured in a dilution refrigerator, using a standard four-wire steady-state method with two RuO$_2$ chip thermometers, calibrated \emph{in situ} against a reference RuO$_2$ thermometer. Magnetic fields were applied along the \emph{c} axis and perpendicular to the heat current. To ensure a homogeneous field distribution in the samples, all fields were applied at temperatures above $T_c$ for transport measurements.

\section{Results and discussion}
\begin{figure}
\includegraphics[width=0.4\textwidth]{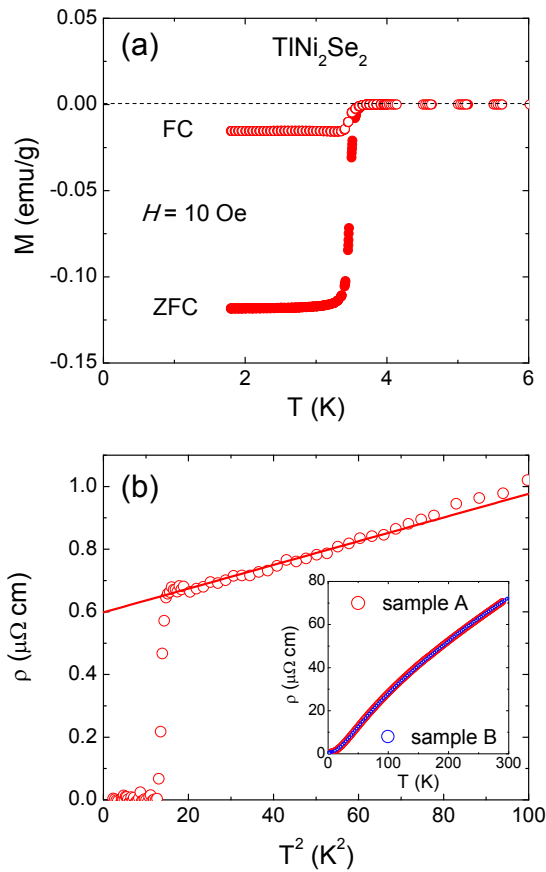}
\caption{(Color online) (a) Low-temperature dc magnetization of TlNi$_2$Se$_2$ single crystal measured with zero-field-cooled (ZFC) and field-cooled (FC) processes. (b) Low-temperature in-plane resistivity of TlNi$_2$Se$_2$ single crystal (sample A). The solid line is a fit of the data between 4 and 8.5 K to $\rho = \rho_0 + AT^2$. Inset shows the $\rho$($T$) curves of sample A and B up to room temperature. After normalizing the value of sample B at 290 K to that of sample A, the two curves are nearly identical.}
\end{figure}

Figure 1(a) shows the low-temperature dc magnetization of TlNi$_2$Se$_2$ single crystal. The onset of the superconducting transition is at 3.7 K. The sharp drop of diamagnetic signal and its quick saturation (below 3.2 K) indicate the sample is of high quality.
Figure 1(b) plots the in-plane resistivity of TlNi$_2$Se$_2$ samples. To reduce the uncertainty associated with geometric factor, we normalize the resistivity of sample B to sample A at $T$ = 290 K. The two resistivity curves are nearly identical after normalization, as seen in the inset. Later we will use the normalized geometric factor for sample B. The resistivity decreases monotonically with lowering the temperature, followed by a sharp superconducting transition. The $T_c$ defined by $\rho = 0$ is 3.7 K, which is consistent with the onset of diamagnetic transition. Fermi-liquid behavior $\rho \sim T^2$ is observed at low temperature. The fit of $\rho(T)$ data between 4 and 8.5 K to $\rho = \rho_0 + AT^2$ gives the residual resistivity $\rho_0$ = 0.60 $\mu\Omega$ cm for sample A and $\rho_0$ = 0.61 $\mu\Omega$ cm for sample B. The residual resistivity ratio (RRR) is about 120, which is much higher than the nickel pnictides BaNi$_2$As$_2$ and SrNi$_2$P$_2$. \cite{Kurita,NKurita}

\begin{figure}
\includegraphics[width=0.447\textwidth]{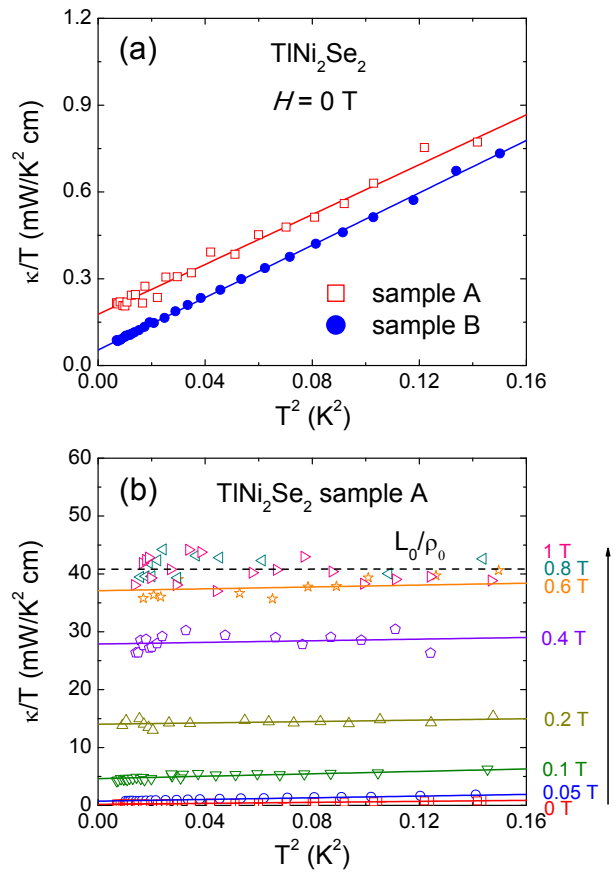}
\caption{(Color online) Low-temperature thermal conductivity of TlNi$_2$Se$_2$ single crystals.
(a) Sample A and sample B in zero field.
(b) Thermal conductivity of sample A under magnetic fields up to $H$ = 1 T. Data below $H_{c2}$ = 0.8 T are fitted to $\kappa/T$ = $\kappa_0/T$ + $bT^2$, as represented by the solid lines.
The dashed line is the normal-state Wiedemann-Franz law expectation $L_0/\rho_0$, with $L_0$ = 2.45 $\times$ 10$^{-8}$ W $\Omega$ K$^{-2}$ and $\rho_0$ = 0.60 $\mu\Omega$ cm.}
\end{figure}

Figure 2(a) shows the low-temperature thermal conductivity of TlNi$_2$Se$_2$ sample A and B at zero field, plotted as $\kappa/T$ versus $T^2$.
The measured thermal conductivity $\kappa$ can be expressed as $\kappa$ = $\kappa_e$ + $\kappa_{ph}$, the sum of electron contribution $\kappa_e$ and
phonon contribution $\kappa_{ph}$. Due to their distinct temperature dependence at low temperatures, the two contributions can be well separated by fitting the data to
\begin{eqnarray}
\kappa = aT + bT^\alpha,
\end{eqnarray}
where $aT$ is the residual linear term of electrons and $bT^{\alpha}$ is the phonon contribution in the boundary scattering limit. Usually $2 < \alpha \leq 3$, which depends on the effect of specular reflection of phonons at the sample surfaces. \cite{Mike,LiSY} For both samples A and B, the fitting parameter $\alpha$ in zero field is very close to 3, therefore we fix it to 3. Note that for BaNi$_2$As$_2$ and SrNi$_2$P$_2$ single crystals, the parameter $\alpha$ is also 3. \cite{Kurita,NKurita} It seems that the effect of specular reflection of phonons at the surfaces is very weak for NBS single crystals.

In Fig. 2(a), the fittings give $\kappa_0/T \equiv a$ = 0.18 $\pm$ 0.02 mW K$^{-2}$ cm$^{-1}$ and 0.05 $\pm$ 0.01 mW K$^{-2}$ cm$^{-1}$ for sample A and B, respectively. Comparing with our experimental error 0.005 mW K$^{-2}$ cm$^{-1}$,\cite{LiSY} these $\kappa_0/T$ values are not negligible. However, they are actually very tiny, if we compare them with the normal-state Wiedemann-Franz law expectation $\kappa_{N0}/T$ = $L_0/\rho_0 \approx$ 40 mW K$^{-2}$ cm$^{-1}$. The ratio ($\kappa_0/T$)/($\kappa_{N0}/T$) of TlNi$_2$Se$_2$ is only 0.44$\%$ (sample A) and 0.12$\%$ (sample B). For nodal superconductors, a substantial $\kappa_0/T$ in zero field contributed by the nodal quasiparticles has been found.\cite{Proust,Suzuki,Hill} For example, $\kappa_0/T$ of the overdoped ($T_c$ = 15 K) $d$-wave cuprate superconductor Tl$_2$Ba$_2$CuO$_{6+\delta}$ (Tl-2201) is 1.41 mW K$^{-2}$ cm$^{-1}$, $\sim$36\% $\kappa_{N0}/T$. \cite{Proust} For the $p$-wave superconductor Sr$_2$RuO$_4$ ($T_c$ = 1.5 K), $\kappa_0/T$ = 17 mW K$^{-2}$ cm$^{-1}$ was reported, more than 9\% $\kappa_{N0}/T$. \cite{Suzuki} The multi-gap nodal heavy-fermion superconductor PrOs$_4$Sb$_{12}$ has $\kappa_0/T$ = 0.46 mW K$^{-2}$ cm$^{-1}$, $\sim$7$\%$ $\kappa_{N0}/T$. \cite{Hill} In this context, the tiny percentage of ($\kappa_0/T$)/($\kappa_{N0}/T$) observed in TlNi$_2$Se$_2$ suggests that the very small $\kappa_0/T$ may not come from nodal quasiparticals.

In fact, the finite value of $\kappa_0/T$ in zero field can be theoretically estimated for a quasi-two-dimensional \emph{d}-wave superconductor: \cite{th1,th2}
\begin{eqnarray}
\frac{\kappa_0}{T}\simeq\frac{\hbar}{2\pi} \frac{\gamma_Nv_F^2}{\triangle_0},
\end{eqnarray}
where $\gamma_N$ is the electronic specific heat coefficient in the normal state, $v_F$ is the Fermi velocity, and $\triangle_0$ stands for the maximum of the superconducting gap.
$\upsilon_F$ = 5.48 $\times$ 10$^4$ m s$^{-1}$, $\gamma_N$ = 40 mJ mol$^{-1}$ K$^{-2}$ and $\triangle_0 = 2.01 k_BT_c$ can be obtained from a former work. \cite{FangMH} In case that TlNi$_2$Se$_2$ is a quasi-two-dimensional \emph{d}-wave superconductor, we estimate $\kappa_0/T$ $\simeq$ 3.22 mW K$^{-2}$ cm$^{-1}$, which should be $\sim$8$\%$ $\kappa_{N0}/T$ of our samples. This value is much higher than what we observed in both sample A and B, therefore, the superconducting gap of TlNi$_2$Se$_2$ is not consistent with the $d$-wave scenario. The very small $\kappa_0/T$ in zero field may result from tiny non-superconducting impure phase in the samples.

\begin{figure}
\includegraphics[width=0.44\textwidth]{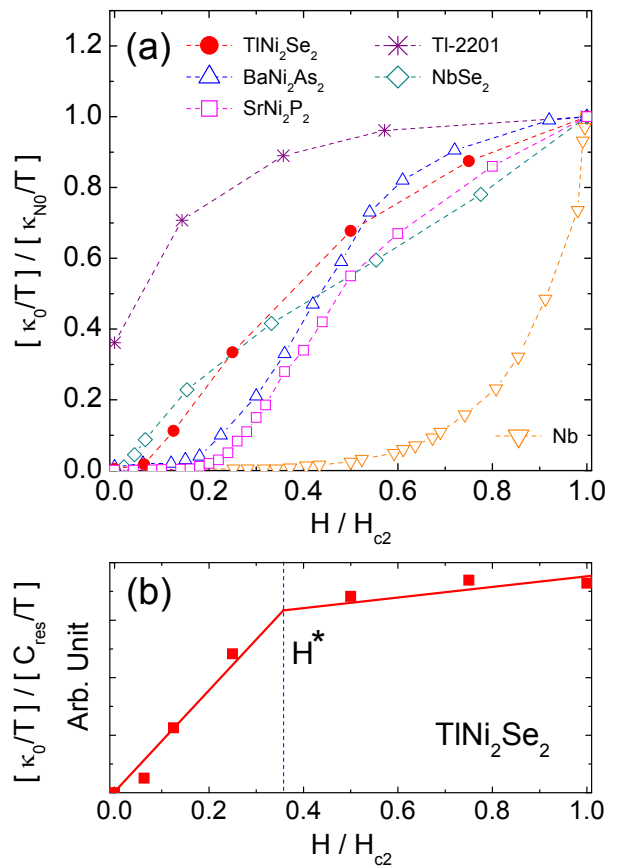}
\caption{(Color online) (a) Normalized $\kappa_0/T$ of TlNi$_2$Se$_2$ as a function of $H/H_{c2}$. For comparison, similar data are shown for the single band $s$-wave superconductor Nb, \cite{Lowell} the multiband $s$-wave superconductor NbSe$_2$, \cite{Boaknin} the $d$-wave superconductor Tl-2201, \cite{Proust} two nickel-pnictide superconductors BaNi$_2$As$_2$ and SrNi$_2$P$_2$. \cite{Kurita,NKurita}
(b) Ratio of thermal conductivity to heat capacity ($\kappa_0/T$)/($C_{res}/T$) in TlNi$_2$Se$_2$. The end of the rapid increase regime at $H^{\ast}$ indicates the complete suppression of the smaller gap(s).}
\end{figure}

The field dependence of $\kappa_0/T$ can provide further information of the superconducting gap structure.\cite{Louis} The thermal conductivity under magnetic fields for sample A is shown in Fig. 2(b). Similar results are obtained for sample B but not shown here, since the data is noisier at $H$ $>$ 0.2 T. Upon applying magnetic fields, vortices are gradually introduced into the sample. The unpaired electrons inside the vortices contribute to $\kappa_0(H)/T$.
Although the curves of 0.8 and 1 T is not smooth, one can still see that $\kappa_0/T$ roughly meets the Wiedemann-Franz law expectation $L_0/\rho_0$ = 40.8 mW K$^{-2}$ cm$^{-1}$. We determine the bulk upper critical field $H_{c2}$ = 0.8 T, which agrees with the value estimated from resistivity measurements. \cite{FangMH} The data in different fields below $H_{c2}$ are also fitted to $\kappa/T = \kappa_0/T + bT^2$, as represented by the solid lines in Fig. 2(b).

Normalized $\kappa_0(H)/T$ of TlNi$_2$Se$_2$ as a function of $H/H_{c2}$ is presented in Fig. 3(a), together with the single band $s$-wave superconductor Nb, \cite{Lowell} the multiband $s$-wave superconductor NbSe$_2$, \cite{Boaknin} the $d$-wave curpate superconductor Tl-2201, \cite{Proust} and two nickel-pnictide superconductors BaNi$_2$As$_2$ and SrNi$_2$P$_2$. \cite{Kurita,NKurita}
For single band $s$-wave superconductor Nb, the $\kappa_0(H)/T$ changes little even up to 40\% $H_{c2}$. \cite{Lowell}
While in nodal superconductor Tl-2201, small field can yield a quick growth in the quasiparticle density of states (DOS) due to Volovik effect, \cite{Volovik} and the low field $\kappa_0(H)/T$ is roughly $\sqrt{H}$ dependant. \cite{Proust}
In the case of NbSe$_2$, the distinct $\kappa_0(H)/T$ behavior was well explained by multiple superconducting gaps with different magnitudes. \cite{Boaknin}

Of the above three archetypal examples, the field dependence of $\kappa_0(H)/T$ for TlNi$_2$Se$_2$ resembles NbSe$_2$ most. Since NBSs also have several bands across the Fermi level, \cite{Terashima,ZhouB,Subedi,RonningF} it is natural to explain the $\kappa_0(H)/T$ behavior of TlNi$_2$Se$_2$ with multiple gaps. As was done in the case of NbSe$_2$, \cite{Boaknin} one can estimate the ratio of smaller gap to larger gap $\triangle_s/\triangle_l$ by plotting the ratio of thermal conductivity to heat capacity ($\kappa_0/T$)/($C_{res}/T$) as a function of $H/H_{c2}$.
In the vortex state, residual specific heat is associated with the unpaired electron DOS, while $\kappa_0/T$ manifests the tunneling ability of those unpaired electrons. Thus ($\kappa_0/T$)/($C_{res}/T$) represents the degree of delocalization of quasiparticles in the vortex state.\cite{Boaknin}
In a simple two-gap model, both Fermi sheets contribute to the rise of ($\kappa_0/T$)/($C_{res}/T$) below a characteristic field $H^{\ast}$. In the regime of $H^{\ast}$ $<$ $H$ $<$ $H_{c2}$, the smaller gap is completely suppressed and only the Fermi sheet with larger gap contributes to the rise of ($\kappa_0/T$)/($C_{res}/T$).

In Fig. 3(b), we plot the ratio ($\kappa_0/T$)/($C_{res}/T$) as a function of $H/H_{c2}$. The residual specific heat $C_{res}(H)/T$ is adopted from Ref. 19.
Two distinct regimes can be resolved: after a rapid initial increase, ($\kappa_0/T$)/($C_{res}/T$) reaches a weak $H$-dependent regime. The end of the rapid increase at $H^{\ast}$ $\simeq$ 0.36 $H_{c2}$ indicates the complete suppression of the smaller gap. Considering that the upper critical field is related to the superconducting gap by $H_{c2}$ $\propto$ $\triangle^2$/$v_F^2$, the characteristic field $H^{\ast}$ allows us to estimate the gap ratio $\triangle_s$/$\triangle_l$ $\simeq$ 0.6. In Ref. 19, the specific heat data can be best fitted by two-gap BCS model with $\triangle_s$/$\triangle_l$ $\simeq$ 0.42, \cite{FangMH} which is qualitatively consistent with our thermal conductivity analysis.

We then compare the $\kappa_0(H)/T$ behavior of TlNi$_2$Se$_2$ to those of BaNi$_2$As$_2$ and SrNi$_2$P$_2$. \cite{Kurita,NKurita} From Fig. 3(a), the three $\kappa_0(H)/T$ curves show a common ``$S$'' shape (concave in low fields and convex in high fields). Previously, this ``$S$''-shape curve of $\kappa_0(H)/T$ was interpreted as the consequence of $H_{c2}$ distribution in the BaNi$_2$As$_2$ and SrNi$_2$P$_2$ crystals due to the sample quality. \cite{Kurita,NKurita}
While for our TlNi$_2$Se$_2$ single crystals, the sharp diamagnetic transition shown in Fig. 1(a) suggests that there should be no $H_{c2}$ distribution, despite the possible existence of tiny impure phase. The RRR of TlNi$_2$Se$_2$ is also much higher than those of BaNi$_2$As$_2$ and SrNi$_2$P$_2$, pointing to cleaner sample. For TlNi$_2$Se$_2$, the Ginzberg-Landau coherence length $\xi$ = 20.3 nm is calculated  from the equation
\begin{eqnarray}
\xi = [\frac{\Upphi_0}{2\pi H_{c2}(0)}]^{\frac{1}{2}},
\end{eqnarray}
where $\Upphi_0$ = 2.07 $\times$ 10$^{-7}$Oe cm$^2$ is the flux quantum.
According to the relationship
\begin{eqnarray}
\frac{\kappa}{T} = \frac{1}{3}\gamma\upsilon_Fl_e,
\end{eqnarray}
the electron mean free path $l_e$ = 677 nm is estimated. The ratio $l_e$/$\xi$ = 33.3 ($\gg$ 1) places our TlNi$_2$Se$_2$ single crystal in the clean limit. Therefore the ``$S$''-shape $\kappa_0(H)/T$ curve of TlNi$_2$Se$_2$ should not be explained by bad sample quality. Since all three compounds have the same crystal structure and similar electronic structure, \cite{RonningF} their common ``$S$''-shape field dependence of $\kappa_0(H)/T$ may have the same origin --- the multiple nodeless gaps.

We note that the normalized $\kappa_0(H)/T$ has been numerically simulated for two band $s$-wave state with unequal gap sizes, which successfully explained the experimental data of Ba(Fe$_{1-x}$Co$_x$)$_2$As$_2$ with different Co-doping by systematically varying the ratio $\triangle_s$/$\triangle_l$. \cite{Bang} However, this kind of calculation can not reproduce the pronounced ``$S$''-shape $\kappa_0(H)/T$ curve in Fig. 3(a), since the simulation is based on an assumption that each band possesses equal weighting of quasiparticle DOS. \cite{Bang} To get the ``$S$''-shape $\kappa_0(H)/T$ curve, one may need to assume that those bands with smaller gap possesses more quasiparticle DOS than those bands with larger gap. Further numerical simulations are needed to reproduce this common feature observed in NBSs.

\section{Conclusion}

In summary, we have measured the thermal conductivity of TlNi$_2$Se$_2$ single crystals down to 50 mK.
The relatively tiny $\kappa_0/T$ and weak field dependance of $\kappa_0(H)/T$ in low fields suggest nodeless superconducting gap.
The $\kappa_0(H)/T$ curve shows an ``$S$''-shape, which was previously also observed in BaNi$_2$As$_2$ and SrNi$_2$P$_2$. This common feature of nickel-based superconductors is explained by multiple nodeless superconducting gaps. A characteristic field $H^{\ast} \simeq$ 0.36 $H_{c2}$ was identified from apparent slope change in ($\kappa_0/T$)/($C_{res}/T$), which gives the ratio $\triangle_s$/$\triangle_l \simeq$ 0.6 in TlNi$_2$Se$_2$.
\\
\\
\begin{center}
{\bf ACKNOWLEDGEMENTS}
\end{center}

This work is supported by the Natural Science Foundation of China (Grants No. 91021016, No. 91221303, No. 11374261, and No. 11204059), the Ministry of Science and Technology of China (National Basic Research Program No. 2012CB821402, No. 2012CB821404, and No. 2011CBA00103), and the Program for Professor of Special Appointment (Eastern Scholar) at Shanghai Institutions of Higher Learning.

\end{document}